\documentclass[12pt,english,floatfix,superscriptaddress,aps,prd,preprint]{revtex4}
\usepackage[utf8]{inputenc}
\usepackage{float}
\usepackage{array}
\usepackage{lipsum}
\usepackage{graphicx}
\usepackage{amsmath,amsthm,amsfonts,amssymb}
\usepackage{graphicx}
\usepackage[english]{babel}
\usepackage{color}
\usepackage{tensor}
\usepackage{esint}
\usepackage[dvips]{epsfig}
\usepackage[dvips]{graphicx}
\usepackage{float}
\usepackage{units}
\usepackage{textcomp}
\usepackage{mathrsfs}
\usepackage{amsmath}
\usepackage[makeroom]{cancel}
\usepackage{amssymb}
\usepackage{amsbsy}
\usepackage{amsfonts}
\usepackage{amssymb,mathrsfs,xcolor}
\usepackage{esint}
\usepackage{array}
\usepackage{graphicx}

\usepackage{hyperref}
\hypersetup{
    colorlinks,
    citecolor=blue,
    filecolor=green,
    linkcolor=purple,
    urlcolor=red,
}

\usepackage{slashed}

\newcommand{\ie}{\begin{equation}}
\newcommand{\fe}{\end{equation}}
\newcommand{\se}{\begin{eqnarray}}
\newcommand{\ff}{\end{eqnarray}}

\usepackage{hyperref}
\hypersetup{colorlinks,breaklinks,
			citecolor=[rgb]{0,0.0,1.0},
            urlcolor=[rgb]{0.0,0.0,1.0},
            linkcolor=[rgb]{0,0.5,0.9}}

\begin{document}

\title{Thermodynamic properties in higher-derivative electrodynamics}


\author{A. A. Ara\'{u}jo Filho}
\email{dilto@fisica.ufc.br}
\affiliation{Universidade Federal do Cear\'a (UFC), Departamento de F\'isica,\\ Campus do Pici, 
Fortaleza - CE, C.P. 6030, 60455-760 - Brazil.}


\author{R. V. Maluf}
\email{r.v.maluf@fisica.ufc.br}
\affiliation{Universidade Federal do Cear\'a (UFC), Departamento de F\'isica,\\ Campus do Pici, Fortaleza - CE, C.P. 6030, 60455-760 - Brazil.}


\date{\today}

\begin{abstract}
In this work, we study the thermodynamic properties of a photon gas in a heat bath within the context of higher-derivative electrodynamics. Specifically, we analyze Podolsky's theory and its extension involving the Lorentz symmetry violation recently proposed in the literature. First, we use the concept of the number of available states of the system in order to construct the partition function. Next, we calculate the main thermodynamic functions: Helmholtz free energy, mean energy, entropy, and heat capacity. In particular, we verify that there exist significant changes in heat capacity and mean energy due to Lorentz violation. Additionally, the modification of the black body radiation and the correction to the \textit{Stefan–Boltzmann} law in the context of the primordial inflationary universe are provided for both theories as well.
\end{abstract}


\maketitle

\section{Introduction}

The concept the mass is a key issue in theoretical physics, particularly within the context of particle physics. For instance, the Higgs mechanism \cite{higgs1964,higgs1966} is the most known approach to generate mass for the particles from a genuine gauge-invariant theory. After all, the interactions between the constituents of matter are usually expressed in terms of gauge theories, which are supposed to be massless. With a different viewpoint, the presence of a massive vector field, commonly ascribed to the Proca’s model, possesses many consequences well encountered in the literature \cite{ryder1996,das2008}. As a result, since the electromagnetic interactions are described in terms of the symmetry group, namely $U(1)$, the Quantum Electrodynamics should be reexamined whether massive modes were taken into account \cite{tu}.

On the other hand, if one deals with the Podolsky electrodynamics \cite{podolsky1942,podolsky1944,podolsky1948}, one will obtain remarkable features. Among them, we can point out that such theory brings about a massive mode without losing the gauge symmetry. Moreover, it was also Podolsky who first attempted to describe the interpretation of this massive mode, i.e.; it was depicted as a neutrino. In this case, the propagator has two poles, one corresponding to the massless mode and the other one associated with the massive mode. Thereby, considering the classical approach, the latter has a feature of removing singularities associated with the pointlike self-energy. Nevertheless, if one regards quantum properties, one will obtain the appearance of ghosts \cite{accioly2010}. The appropriate gauge condition to provide Podolsky electrodynamics is no longer the common Lorenz gauge but rather a modified one \cite{pimentel}, being consistent with the existence of five degrees of freedom. Two of them are related to the massless photon mode while the other three ones are related to the massive longitudinal mode \cite{casana2018}. Besides, it is worth mentioning that, in the presence of the Podolsky term, there exist other remarks involving quantum field theory in the context of renormalization \cite{bufalo2012}, path integral quantization and fine-temperature approach \cite{bufalo2011,gaete,bonin2010}, multipole expansions \cite{bonin2019}, black holes \cite{cuzinatto2018}, cosmology \cite{cuzinatto2017}  and others \cite{kruglov2010,cuzinatto2011,zayats2014,granado2019,nogueira2019,borges2019}.

About twenty years up to now, the Lorentz-violating contributions of mass dimensions 3 and 4 have been taken into account within both theoretical and phenomenological scenarios in the photon and lepton sectors \cite{bonetti2017}. Recently, theories with higher-dimension operators have received much attention after a generalized approach proposed by Mewes and Kostelecký in Ref. \cite{kostelecky2013}. In the CPT-even photon sector, the leading-order contributions in an expansion in terms of additional derivatives are dimension-6 ones. Hence, these are also the most prominent ones that could play a role in nature if higher-derivative Lorentz violation (LV) existed \cite{casana2018}. In such a way, there is a lack in the literature concerning the study of its respective thermodynamic properties. In this context, it is important to conduct further investigations. Therefore, the physical implications of the thermodynamic properties of such theories should be taken into account in order to possibly address some fingerprints of a new physics that might be mapped into future applications in either condensed matter physics or statistical mechanics. Thereby, we present a theoretical background in order to might serve as a basis for further experimental studies seeking any trace of Lorentz violation.

In this sense, this work provides a study in such direction, highlighting the behavior of the main thermodynamic functions such as Helmholtz free energy, mean energy, entropy, and heat capacity. Furthermore, the corrections to the black body radiation and the \textit{Stefan–Boltzmann} law are analyzed as well.


\section{Podolsky electrodynamics}

\subsection{The model}

The four-dimensional Lagrangian density of the Podolsky electrodynamics is written as
\cite{podolsky1942,podolsky1944,podolsky1948}  
\ie
\mathcal{L}= - \frac{1}{4}F_{\mu\nu}F^{\mu\nu} + \frac{\theta^{2}}{2}\partial_{\alpha}F^{\alpha\beta}\partial_{\lambda}F\indices{^\lambda_\beta} -A_{\mu}J^{\mu}, \label{1}
\fe
where $F_{\mu\nu}= \partial_{\mu}A_{\nu} - \partial_{\nu}A_{\mu}$ is the usual Maxwell field strength, $A_{\mu}$ is the vector field and $\theta$ is the Podolsky's parameter with mass dimension -1, and $J^{\mu}$ is a conserved current. In addition, in Refs. \cite{pimentel,gaete} the authors accomplished remarkable classical analyzes of this theory i.e., they studied interparticle potential between sources, quantization, generalization of such theory in the framework of Dirac's theory of constrained systems and others. From Eq. (\ref{1}), the equation of motion can be written as 
\ie
\left( 1+ \theta^{2} \Box  \right)\partial_{\mu}F^{\mu\nu}= J^{\nu}.
\fe
Here, it must be pointed out that there exists a distinguishing characteristic if compared with the Proca's theory which is the generation of a massive mode without losing its gauge symmetry. Thereby, by adding a gauge fixing term to Eq. (\ref{1}), namely $-\frac{1}{2\varsigma}(\partial_{\mu} A^{\mu})^{2}$, with $\varsigma$ being the gauge-fixing parameter, we can immediately derive the propagator in the momentum space as follows
\ie
\Delta_{\mu\nu}(k)= - \frac{1}{k^{2}(1-\theta^{2}k^{2})} \bigg\{ \theta_{\mu\nu}+\varsigma(1-\theta^{2}k^{2})\omega_{\mu\nu} \bigg\},\label{prop1}
\fe
where $\theta_{\mu\nu}\equiv\eta_{\mu\nu}-\omega_{\mu\nu}$ and $\omega_{\mu\nu}\equiv k_{\mu}k_{\nu}/k^2$ are the transverse and longitudinal projectors respectively. Here, $\eta_{\mu\nu}$ is the Minkowski metric with signature $\left(+.-,-,-\right)$. Clearly, from above expression, we verify the presence of both Maxwell $k^{2}=0$ and Podolsky $1-\theta^{2}k^{2}=0$ poles. From the pole of the propagator encountered in Eq. (\ref{prop1}), we have
\ie
k_{\mu}k^{\mu}(1-\theta^{2} k^{\alpha}k_{\alpha})=0.
\label{dispe}
\fe
Furthermore, note that when one considers $\theta \rightarrow 0$, the standard dispersion relation is recovered i.e., $k^{2}=0$. Besides, Eq. (\ref{dispe}) may be rewritten simply as
\ie
{\bf{k}}^{2} = \frac{2E^{2}\theta^{2} -1 \mp 1}{2\theta^{2}} \label{dr},
\fe
where this equation shows that we have a different state equation which must change the thermodynamic properties of our system due to the fact that the relation between energy and momentum is no longer ascribed to be the usual one. It is worth to note that from Eq. (\ref{dr}), we choose the -1 configuration, since otherwise we will not have the contribution of parameter $\theta$. Moreover, in what follows, we examine a photon gas in a volume $\Gamma$ and instead of dealing with a quantizing momentum due to the boundary conditions, rather we assume a continuous momentum spectrum as it is commonly used in the literature \cite{reif,camacho2007,anacleto2018}. We use the fact that the statistical mechanics tells us that the relation between energy and momentum has a remarkable aspect in evaluating the dependence of the pressure as a function of the energy density. 
In the next subsection, we proceed with the purpose of obtaining the accessible states of the system in order to calculate the partition function which suffices to address all thermodynamic properties. In addition, it is noteworthy that in different contexts the thermodynamic functions were calculated as well \cite{oliveira2019,oliveira2020,hassanabadi2016,pacheco2014,yao2018}.

\subsection{Thermodynamic properties}

\begin{figure}[ht]
\centering
\includegraphics[scale=0.4]{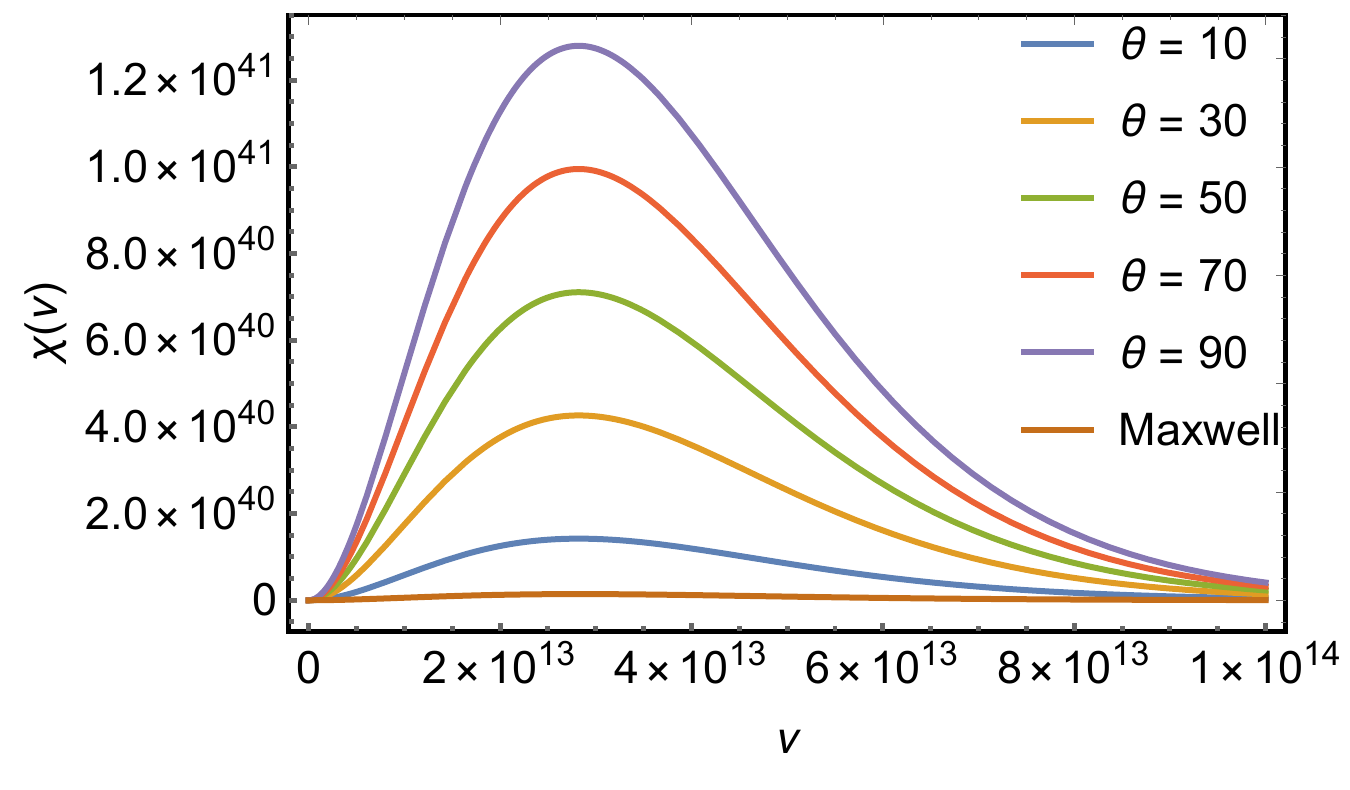}
\caption{The figure shows the behavior of the black body radiation as a function of the frequency $\nu$ for different values of $\theta$ within Podolsky electrodynamics considering $h=1$.}
\label{p}
\end{figure}

We start with the construction of the partition function for the sake of obtaining the following thermodynamic properties i.e., Helmholtz free energy, mean energy, entropy heat capacity. In this sense, we use the traditional method for doing so; we use the concept of the number of accessible states of the system \cite{reif}. Generically, it can be written as
\ie
\Omega(E) = \frac{\zeta}{(2\pi)^{3}}\int \int \mathrm{d}^{3} {\bf{x}} \,\mathrm{d}^{3} {\bf{k}},
\fe
where $\zeta$ is the spin multiplicity which in our case will be considered as the photon sector i.e., $\zeta = 2$. However, for the sake of simplicity, the above equation may be rewritten as follows
\ie
\Omega(E) = \frac{\Gamma}{\pi^{2}} \int^{\infty}_{0} \mathrm{d} {\bf{k}} |{\bf{k}}|^{2}, \label{ms2}
\fe
where $\Gamma$ is considered the volume of the thermal bath and $\mathrm{d} {\bf{k}}$ being given by
\ie
\mathrm{d} {\bf{k}} = \frac{\theta E}{\sqrt{E^{2}\theta^{2}-1}} \mathrm{d}E.
\label{vol}
\fe
After substituting $(\ref{dr})$ and $(\ref{vol})$ in $(\ref{ms2})$, we obtain
\ie
\Omega(E) = \frac{\Gamma}{\pi^{2}} \int^{\infty}_{0}  \frac{E(E^{2} \theta^{2}-1)} {\theta\sqrt{E^{2}\theta^{2}-1}} \,\mathrm{d}E,
\fe
and therefore we are properly able to write down the partition function analogously to what the authors did in Ref. \cite{anacleto2018} as follows
\ie
\mathrm{ln}\left[ Z(\beta,\Gamma)\right] = - \frac{\Gamma}{\pi^{2}} \int^{\infty}_{0}  \frac{E(E^{2} \theta^{2}-1)} {\theta\sqrt{E^{2}\theta^{2}-1}} \mathrm{ln} \left(  1- e^{-\beta E} \right)\mathrm{d}E,
\label{partition}
\fe
where $\beta = 1/k_{B}T$. Using Eq.(\ref{partition}), we can obtain the thermodynamic functions per volume $\Gamma$, namely, Helmholtz free energy $F(\beta,\theta)$, mean energy $U(\beta,\theta)$, entropy $S(\beta,\theta)$ and heat capacity $C_{V}(\beta,\theta)$ defined as follows:
\ie
\begin{split}
 & F(\beta,\theta)=-\frac{1}{\beta} \mathrm{ln}\left[Z(\beta,\theta)\right], \\
 & U(\beta,\theta)=-\frac{\partial}{\partial\beta} \mathrm{ln}\left[Z(\beta,\theta)\right], \\
 & S(\beta,\theta)=k_B\beta^2\frac{\partial}{\partial\beta}F(\beta,\theta), \\
 & C_V(\beta,\theta)=-k_B\beta^2\frac{\partial}{\partial\beta}U(\beta,\theta).
\label{properties}
\end{split}
\fe  
At the beginning, let us consider the mean energy
\ie
U(\beta,\theta) = \frac{1}{\pi^{2}}  \int^{\infty}_{0}  \frac{E^{2}\sqrt{E^{2} \theta^{2}-1}\,e^{-\beta E}} {\theta\left(  1- e^{-\beta E} \right)} \mathrm{d}E, \label{meanenergy}
\fe
which follows the spectral radiance defined by:
\ie
\chi(\theta,\nu) = \frac{(h\nu)^{2}\sqrt{(h\nu)^{2} \theta^{2}-1}\,e^{-\beta h\nu}} {\pi^{2}\theta\left(1- e^{-\beta h\nu} \right)},
\label{spectralradiance}
\fe
with $E=h\nu$ where $h$ is the Planck constant and $\nu$ is the frequency. Here, it is reasonable to investigate how the parameter $\theta$ affects our theory in the spectral radiation. Additionally, it has to be noted that, even though we explicit the constants $h,k_{B}$, for performing the following calculations, we set them $h=k_{B}=1$.  In this way, the plot of this configuration is shown in Fig. \ref{p}. Here, we notice that the black body radiation spectra for different values of $\theta$ are greater than one exhibited in the Maxwell case. On the other hand, when $\theta \rightarrow 0$, we recover the usual Maxwell electrodynamics. Physically, such result reflects the existence of an additional massive mode presented in Podolsky electrodynamics.

\begin{figure}[ht]
\centering
\includegraphics[width=6cm,height=4cm]{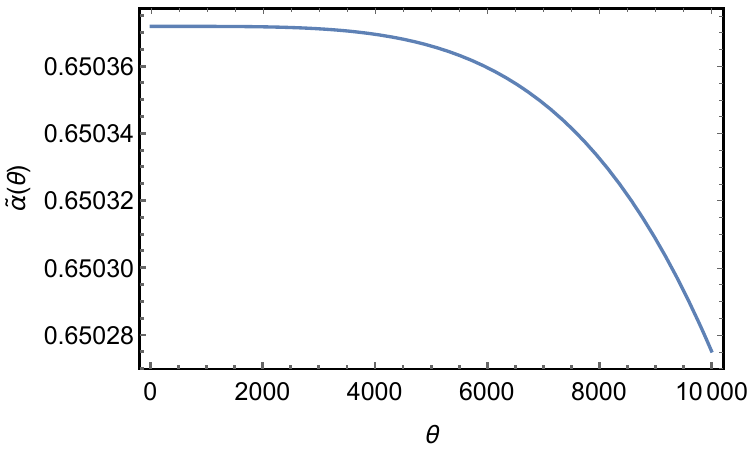}
\includegraphics[width=6cm,height=4cm]{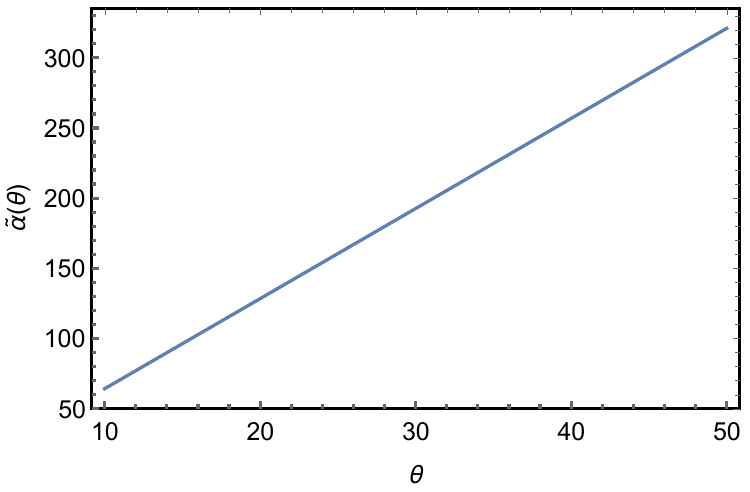}
\includegraphics[width=6cm,height=4cm]{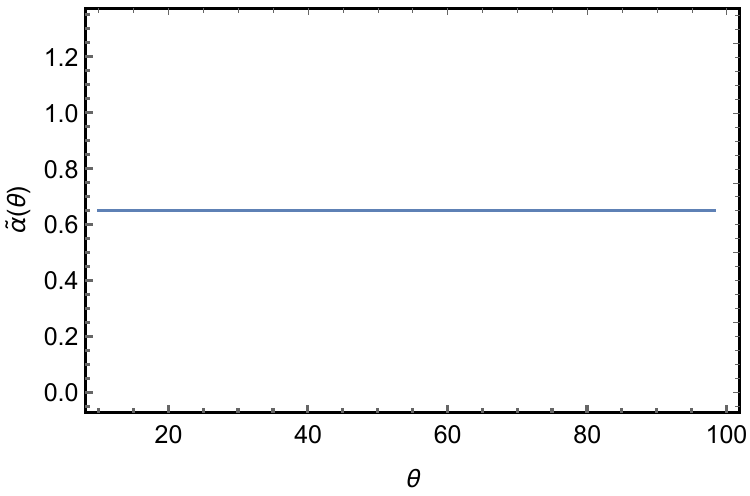}
\caption{The figure shows the correction to the \textit{Stefan–Boltzmann} law represented by parameter $\tilde{\alpha}$ as a function of $\theta$ considering $k_{B}=1$ and the temperature in the early inflationary universe i.e., $\beta = 10^{-13}$ GeV.}
\label{alphas}
\end{figure}

For the sake of obtaining the well-established radiation constant of the \textit{Stefan- Boltzmann} energy density i.e., $u_{S}= \alpha T^{4}$, we consider $(E \theta)^{2} \gg 1$ which leads to
\ie
\alpha = \frac{1}{\pi^{2}}  \int^{\infty}_{0}  \frac{E^{3}\,e^{-\beta E}} {\left(  1- e^{-\beta E} \right)} \mathrm{d}E = \frac{\pi^{2}}{15},
\label{radiance}
\fe
reproducing the well-established result in the literature \cite{zettili2003}. On the other hand, in order to check how the coupling constant $\theta$ affects the new radiation constant, we proceed as follows:
\ie
\tilde{\alpha} \equiv U(\beta,\theta) \beta^{4}. \label{sbl}
\fe
The analysis will be accomplished via numerical calculations. The plots are shown in Fig.\ref{alphas} taking into account three different circumstances i.e., when $\theta$ is either a small or a huge number. Furthermore, the aspect of examining the limit when $E^{2}\sqrt{E^{2} \theta^{2}-1} \gg \theta$ is also regarded. The latter can be handily associated with the
primordial inflationary universe, since we may deal with high temperature regime i.e., $\beta= 10^{-13}$ GeV. Another interesting approach which it is worth being investigated is whether the thermodynamics functions bear with CMB (Cosmic Microwave Background) analysis. Nevertheless, this approach lies beyond the scope of the current work and will be addressed in a future upcoming manuscript though.

\begin{figure}[ht]
\centering
\includegraphics[width=6cm,height=4cm]{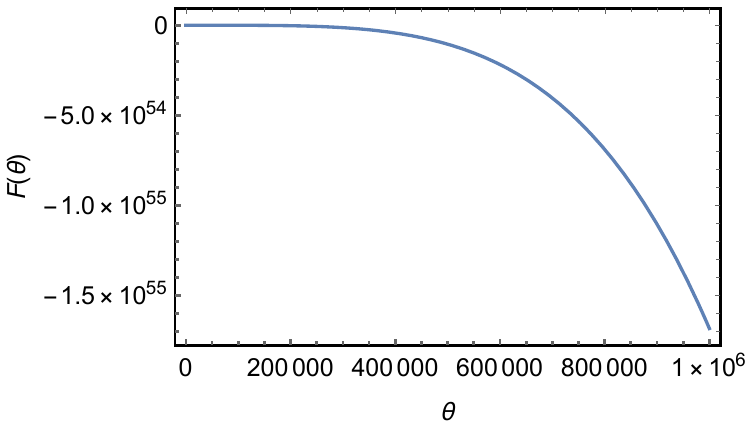}
\includegraphics[width=6cm,height=4cm]{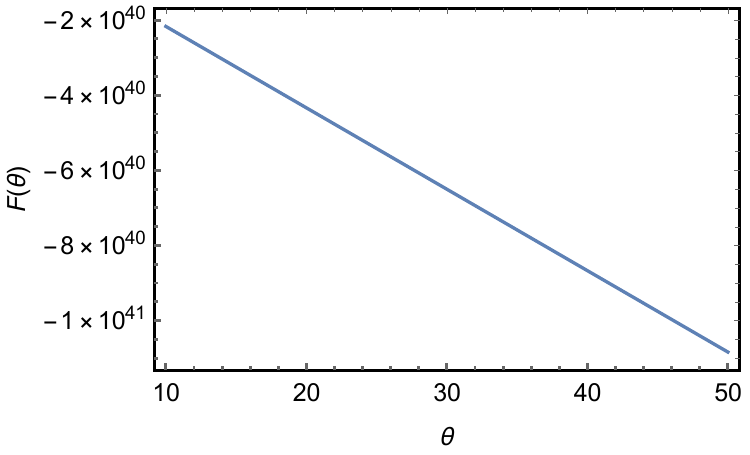}
\includegraphics[width=6cm,height=4cm]{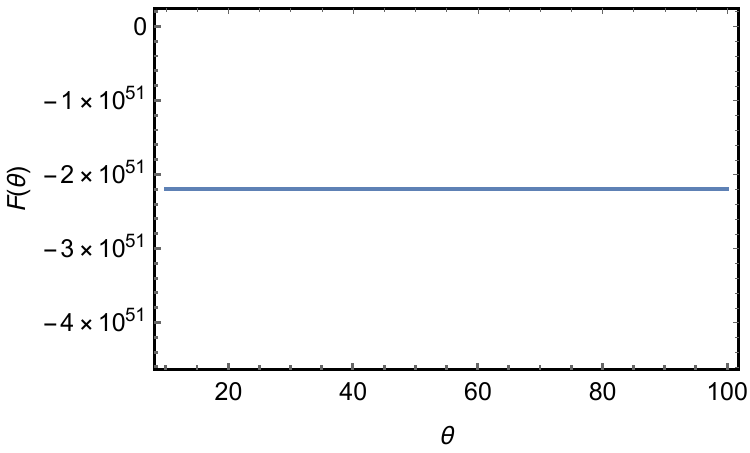}
\caption{The figure exhibits the solutions to Helmholtz free energy $F(\theta)$ at high temperature regime considering $k_{B}=1$.}
\label{fs}
\end{figure}

\begin{figure}[ht]
\centering
\includegraphics[width=6cm,height=4cm]{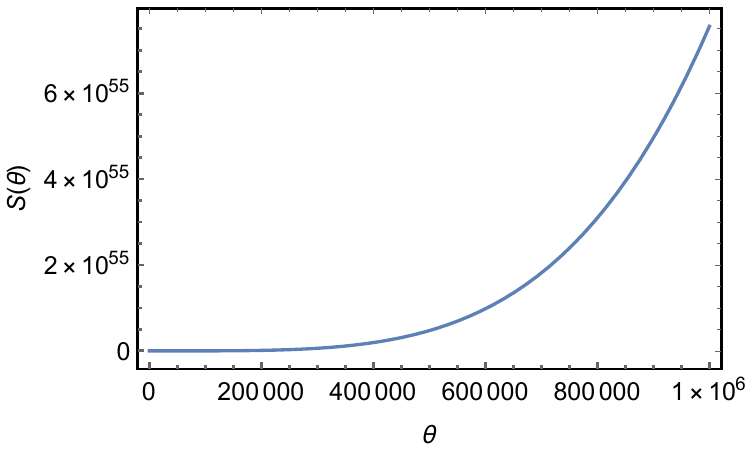}
\includegraphics[width=6cm,height=4cm]{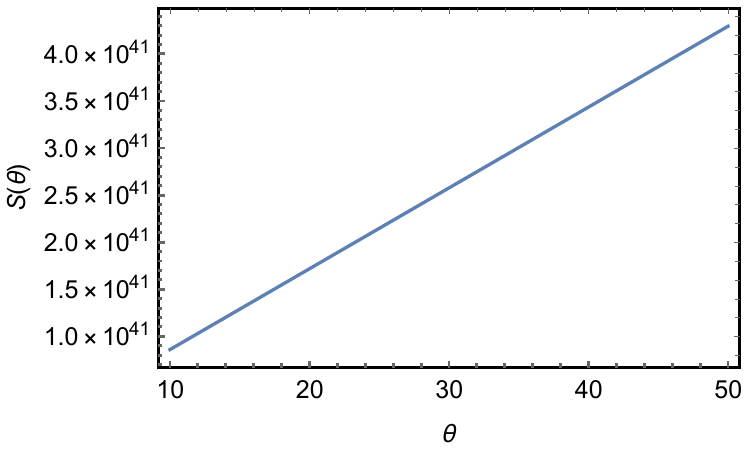}
\includegraphics[width=6cm,height=4cm]{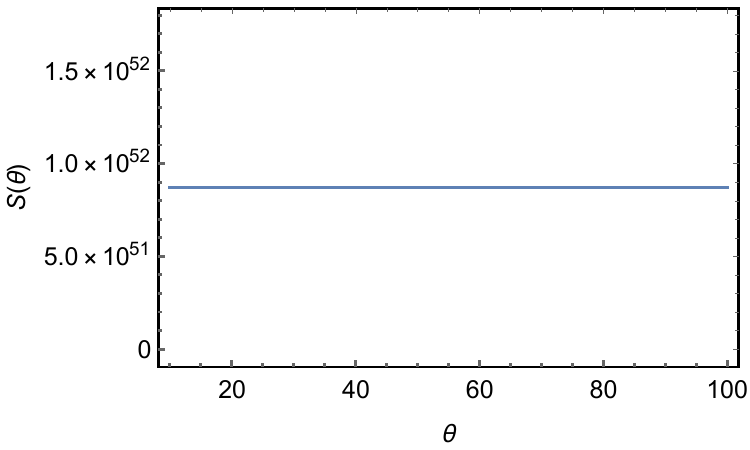}
\caption{These three graphics exhibit the behavior of the entropy $S(\theta)$ considering the temperature in the primitive inflationary universe where $k_{B}=1$.}
\label{ss}
\end{figure}

\begin{figure}[ht]
\centering
\includegraphics[width=6cm,height=4cm]{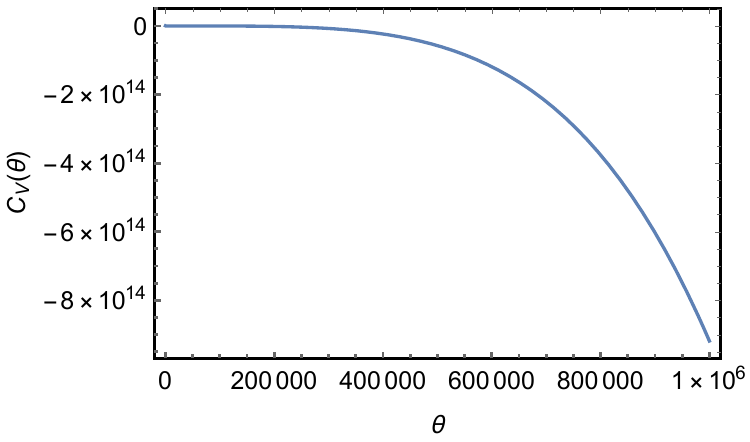}
\includegraphics[width=6cm,height=4cm]{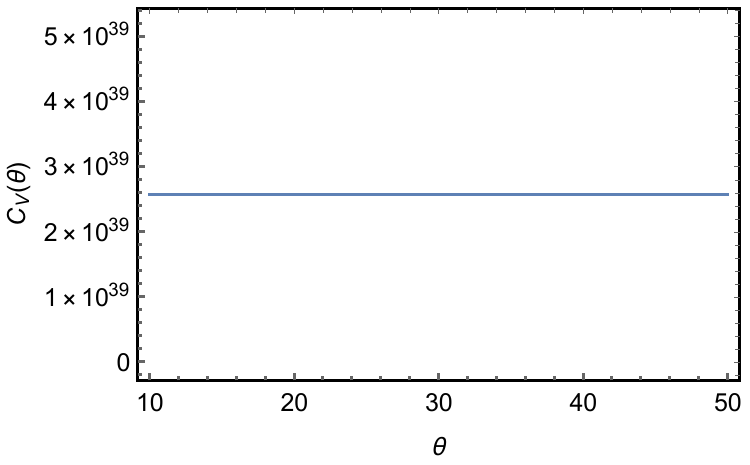}
\includegraphics[width=6cm,height=4cm]{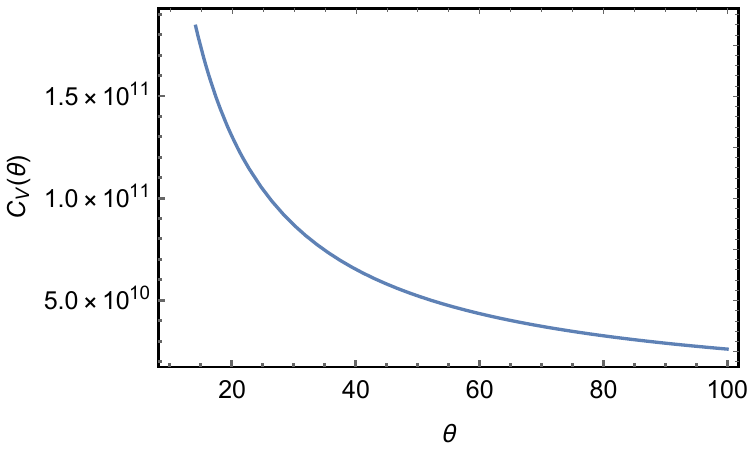}
\caption{The plots exhibit solutions to the heat capacity $C_{V}(\theta)$ considering $k_{B}=1$ and the temperature at the beginning inflationary universe.}
\label{cc1}
\end{figure}

Analogously, the remaining thermodynamic functions can be explicitly computed:
\ie
F(\beta,\theta) = \frac{1}{\theta \pi^{2} \beta} \int^{\infty}_{0} E \sqrt{E^{2}\theta^{2}-1}\,\mathrm{ln}\left( 1-e^{-\beta E}\right) \mathrm{d}E, \label{helmontz}
\fe
\ie
S(\beta,\theta) = \frac{k_{B}}{\theta \pi^{2}} \left( -\int^{\infty}_{0} E \sqrt{ E^{2}\theta^{2}-1}\,\mathrm{ln}\left( 1-e^{-\beta E}\right)+ \beta\int^{\infty}_{0} \frac{E^{2} \sqrt{ E^{2}\theta^{2}-1} \,e^{-\beta E}}{1-e^{-\beta E}}\right) \mathrm{d}E, \label{entropy}
\fe
\ie
C_{V}(\beta,\theta) = \frac{ k_{B} \beta^{2}}{ \theta \pi^{2}} \left( +\int^{\infty}_{0} \frac{E^{3} \sqrt{ E^{2}\theta^{2}-1}\, e^{-2 \beta E}}{\left(1- e^{-\beta E}\right)^{2}}+ \int^{\infty}_{0} \frac{E^{3} \sqrt{ E^{2}\theta^{2}-1} \,e^{-\beta E}}{1-e^{-\beta E}}\right) \mathrm{d}E, \label{heatcapacity}
\fe
and the following results ascribed to them are displayed in Figs. \ref{fs}, \ref{ss} and \ref{cc1} respectively.

\section{Podolsky with Lorentz violation}

\subsection{The model}

Here, we study the Podolsky electrodynamics modified by the
traceless LV dimension-6 term presented in Ref.\cite{casana2018}. In this work, the authors analyzed the classical aspects of such theory taking into account unitariry and causality from the study of the respective propagator proceeding analogously to it is already established in the literature \cite{maluf2019,scarpelli2003}. They consider the construction of a closed algebra using the prescription $D_{\mu\nu}=(B_{\mu}C_{\nu}-B_{\nu}C_{\mu})/2$, where $B_{\mu}$ and $C_{\nu}$ are constant background vectors which accounts for LV. In this sense, the Lagrangian density which represents this model is given by
\ie
\mathcal{L}= - \frac{1}{4}F_{\mu\nu}F^{\mu\nu} + \frac{\theta^{2}}{2}\partial_{\alpha}F^{\alpha\beta}\partial_{\lambda}F\indices{^\lambda_\beta} +\eta^{2}D_{\beta\alpha}\partial_{\sigma}F^{\sigma\beta}\partial_{\lambda}F^{\lambda\alpha} + \frac{1}{2\xi}(\partial_{\mu}A^{\mu})^{2}, \label{LV}
\fe
where $\theta$ is the same parameter defined previously, $\eta$ is the constant coupling with dimension of mass $[m]^{-1}$ and $\xi$ is the gauge fixing parameter required to evaluate the respective propagator. Nevertheless, we focus only on the investigation of its dispersion relation presented in the poles of the propagator for the sake of calculating the following thermodynamic functions. Therefore, the poles are given by
\ie
k^{2}(1-\theta^{2}k^{2})\gamma(k) =0, \label{dr2}
\fe
where $\gamma(k)=\eta^{4}[(B \cdot k)^{2}-B^{2}k^{2}][(C \cdot k)^{2}- C^{2}k^{2}] - \left[1- \theta^{2}k^{2}-\eta^{2}(B \cdot C)k^{2}+\eta^{2}(B \cdot k)(C \cdot k)\right]^{2}$. Considering the complete isotropic sector in this theory, i.e., $D_{\mu\nu}=-D_{00} \times \mathrm{diag}(3,1,1,1)_{\mu\nu}$, Eq. (\ref{dr2}) turns out to be written as 
\ie
E = \sqrt{\left( 1- \frac{8 \eta^{2}D_{00}}{\theta^{2}+2\eta^{2}D_{00}}  \right){\bf{k}}^{2} + \frac{1}{\theta^{2}+2\eta^{2}D_{00}}}.
\fe
As it was accomplished in the last section, in what follows, we calculate the accessible states for this configuration which accounts for the Lorentz violation. Next, we proceed likewise.


\subsection{Thermodynamic properties}

\begin{figure}[ht]
\centering
\includegraphics[scale=0.6]{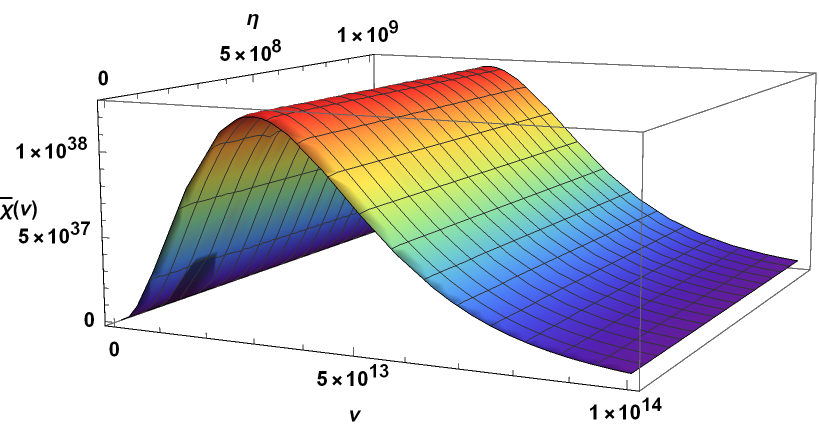}
\caption{The figure shows the behavior of the black body radiation in the generalized Podolsky electrodynamics with Lorentz violation for fixed values of $\theta=10$ and $D_{00}=1$.}
\label{lv}
\end{figure}

The number of accessible states per volume is  
\ie
\Bar{\Omega}(E) = \frac{1}{\pi^{2}} \int^{\infty}_{0}  E\left(1- \frac{8\eta^{2}D_{00}}{\theta^{2} +2\eta^{2}D_{00}}\right)^{-3/2} \left[ \frac{E^{2} (\theta^{2} +2\eta^{2}D_{00})-1}{{\theta^{2} +2\eta^{2}D_{00}}}\right]^{1/2}  \,\mathrm{d}E,
\fe
and, from it, we can construct the respective partition function for such theory which follows 
\ie
\mathrm{ln}[\Bar{Z}(\beta,\eta,\theta,D_{00})] = -\frac{1}{\pi^{2}} \int^{\infty}_{0}  E\left(1- \frac{8\eta^{2}D_{00}}{\theta^{2} +2\eta^{2}D_{00}}\right)^{-3/2} \left[ \frac{E^{2} (\theta^{2} +2\eta^{2}D_{00})-1}{{\theta^{2} +2\eta^{2}D_{00}}}\right]^{1/2} \mathrm{ln} \left( 1 - e^{-\beta E} \right) \,\mathrm{d}E
\fe
 and using the definitions established in (\ref{properties}), we calculate Helmholtz free energy, mean energy, entropy, and heat  capacity. At the beginning, we devote our attention to the spectral radiance as we did before. The respective plot of $\bar{\chi}$ as a function of frequency $\nu$ for different values of $\eta$ is shown in Fig. \ref{lv}. In agreement with the previous section, in which we accomplished the correction to the \textit{Stefan–Boltzmann} law exhibited in Eq. (\ref{sbl}), we step forward likewise
\ie
\Bar{\alpha} \equiv \Bar{U}(\beta,\eta,\theta,D_{00}) \beta^{4}. \label{sbl2}
\fe 
 Again, this analysis will be performed via numerical approach in the context of primordial temperature of the universe. To perform such calculation, we need to obtain the behavior of the mean energy. In this way, 
\ie
\Bar{U}(\beta,\eta,\theta,D_{00}) = \frac{1}{\pi^{2}} \int^{\infty}_{0}  E^{2}\left(1- \frac{8\eta^{2}D_{00}}{\theta^{2} +2\eta^{2}D_{00}}\right)^{-3/2} \left[ \frac{E^{2} (\theta^{2} +2\eta^{2}D_{00})-1}{{\theta^{2} +2\eta^{2}D_{00}}}\right]^{1/2} \frac{e^{-\beta E}}{1-e^{-\beta E}} \,\mathrm{d}E, \label{meanenergy2}
\fe
with the spectral radiance (plotted in Fig. \ref{lv}) being given by 
\ie
\Bar{\chi}(\beta,\eta,\theta,D_{00}) = \frac{1}{\pi^{2}}   E^{2}\left(1- \frac{8\eta^{2}D_{00}}{\theta^{2} +2\eta^{2}D_{00}}\right)^{-3/2} \left[ \frac{E^{2} (\theta^{2} +2\eta^{2}D_{00})-1}{{\theta^{2} +2\eta^{2}D_{00}}}\right]^{1/2} \frac{e^{-\beta E}}{1-e^{-\beta E}}.
\fe
Here, the remaining thermodynamics functions are provided bellow:
the Helmholtz free energy

\ie
\Bar{F}(\beta,\eta,\theta,D_{00}) =
\frac{1}{\pi^{2}\beta^{2}} \int^{\infty}_{0}  E\left(1- \frac{8\eta^{2}D_{00}}{\theta^{2} +2\eta^{2}D_{00}}\right)^{-3/2} \left[ \frac{E^{2} (\theta^{2} +2\eta^{2}D_{00})-1}{{\theta^{2} +2\eta^{2}D_{00}}}\right]^{1/2} \mathrm{ln} \left( 1 - e^{-\beta E} \right) \,\mathrm{d}E, \label{helmontz2}
\fe
the entropy
\ie
\begin{split}
\Bar{S}(\beta,\eta,\theta,D_{00}) = &-
\frac{1}{\pi^{2}} \int^{\infty}_{0} E \left(1- \frac{8\eta^{2}D_{00}}{\theta^{2} +2\eta^{2}D_{00}}\right)^{-3/2} \left[ \frac{E^{2} (\theta^{2} +2\eta^{2}D_{00})-1}{{\theta^{2} +2\eta^{2}D_{00}}}\right]^{1/2} \mathrm{ln} \left( 1 - e^{-\beta E} \right) \,\mathrm{d}E\\
& + \frac{\beta}{\pi^{2}} \int^{\infty}_{0} E^{2} \left(1- \frac{8\eta^{2}D_{00}}{\theta^{2} +2\eta^{2}D_{00}}\right)^{-3/2} \left[ \frac{E^{2} (\theta^{2} +2\eta^{2}D_{00})-1}{{\theta^{2} +2\eta^{2}D_{00}}}\right]^{1/2} \frac{e^{-\beta E}}{1-e^{-\beta E}} \mathrm{d}E, \label{entropy2}
\end{split}
\fe
and finally, the heat capacity
\ie
\begin{split}
\Bar{C}_{V}(\beta,\eta,\theta,D_{00}) = &+ \frac{\beta^{2}}{\pi^{2}} \int^{\infty}_{0} E^{3} \left(1- \frac{8\eta^{2}D_{00}}{\theta^{2} +2\eta^{2}D_{00}}\right)^{-3/2} \left[ \frac{E^{2} (\theta^{2} +2\eta^{2}D_{00})-1}{{\theta^{2} +2\eta^{2}D_{00}}}\right]^{1/2} \frac{ e^{-2\beta E}}{\left(1-e^{-\beta E}\right)^{2}}\mathrm{d}E \\
&+ \frac{\beta^{2}}{\pi^{2}} \int^{\infty}_{0} E^{3} \left(1- \frac{8\eta^{2}D_{00}}{\theta^{2} +2\eta^{2}D_{00}}\right)^{-3/2} \left[ \frac{E^{2} (\theta^{2} +2\eta^{2}D_{00})-1}{{\theta^{2} +2\eta^{2}D_{00}}}\right]^{1/2} \frac{e^{-\beta E}}{1-e^{-\beta E}} \mathrm{d}E. \label{heatcapacity2}
\end{split}
\fe

\begin{figure}[ht]
\centering
\includegraphics[width=6cm,height=4cm]{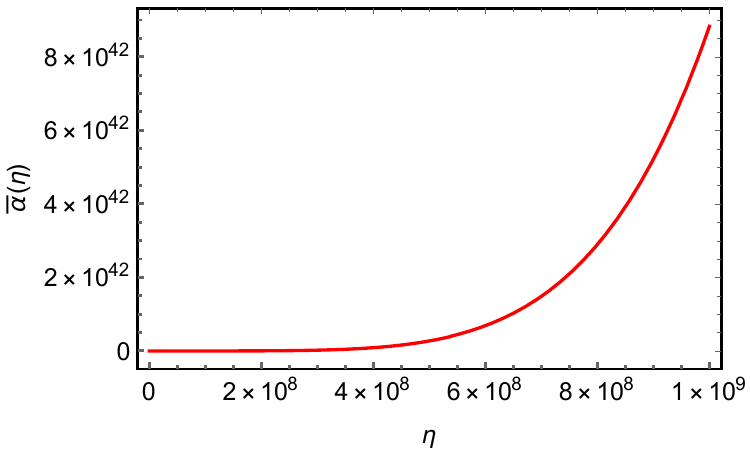}
\includegraphics[width=6cm,height=4cm]{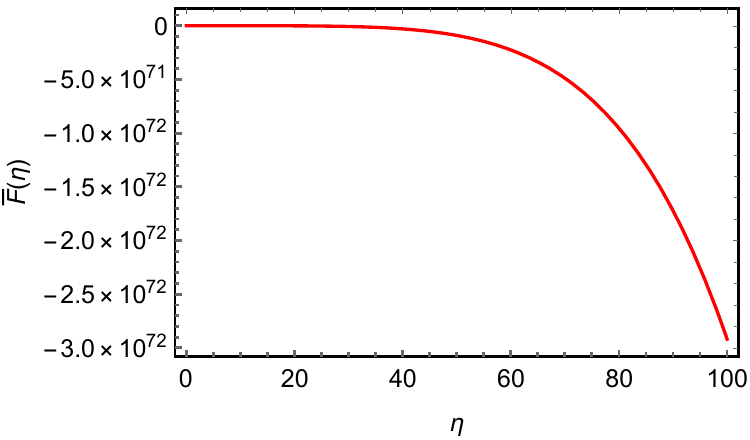}
\includegraphics[width=6cm,height=4cm]{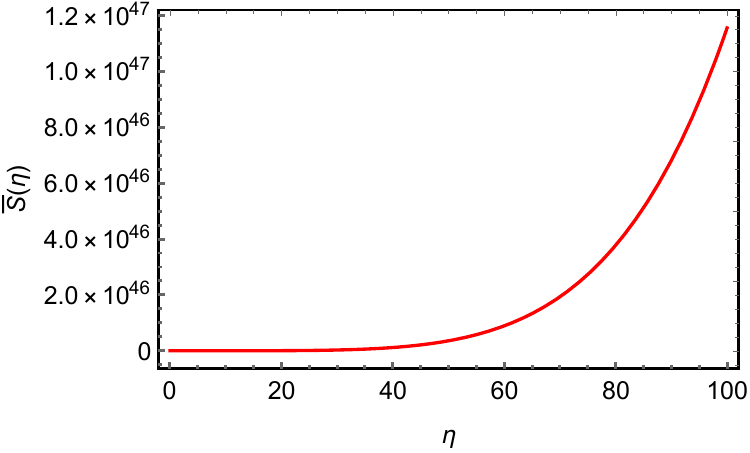}
\includegraphics[width=6cm,height=4cm]{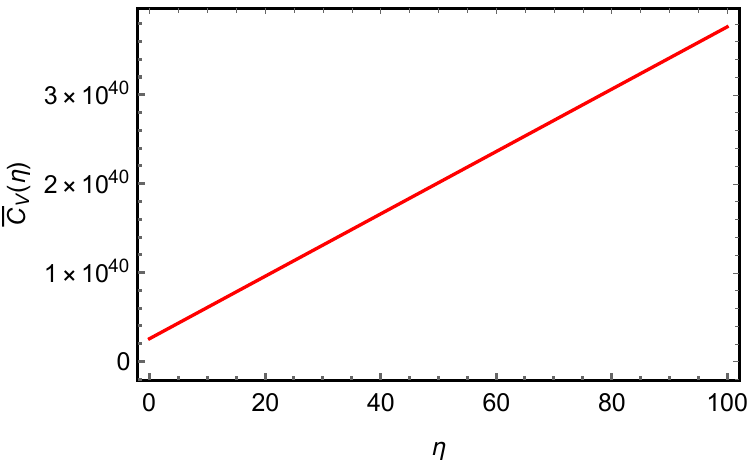}
\caption{The graphics show the correction to the \textit{Stefan–Boltzmann} law $\Bar{\alpha}(\eta)$, entropy $\Bar{S}(\eta)$, Helmholtz free energy $\bar{F}(\eta)$ and heat capacity $\Bar{C}_{V}(\eta)$ regarding $k_{B}=1$.}
\label{tlv}
\end{figure}


\section{Results and discussions}

Initially, let us focus our attention on the pure Podolsky electrodynamics. Indeed, we determined the expression of the spectral radiance $\chi (\nu)$ exhibited in (\ref{spectralradiance}) with its plot displayed in Fig. \ref{p}. From it, we could see that $\chi (\nu)$ was sensitive to changes of $\theta$ i.e., $\theta$= 10, $\theta$= 30, $\theta$= 50, $\theta$= 70, $\theta$= 90. We made a comparison of these different values to the black body radiation with the Maxwell theory. We noted that the latter had its spectral radiance smaller than the Podolsky one. It is important to note that to accomplish such analysis, we needed to consider the limit where $E^{2}\sqrt{E^{2} \theta^{2}-1} \gg \theta$. Next, we did the plot of Fig. \ref{alphas} which showed the correction to the \textit{Stefan–Boltzmann} law represented by parameter $\tilde{\alpha}(\theta)$ considering the temperature in the early inflationary universe i.e., $\beta = 10^{-13}$ GeV. The upper graph in the left side exhibited a very slow variation of $\tilde{\alpha}(\theta)$ when $\theta$ varied. Moreover, we considered small variations of $\theta$, which entailed a constant behavior (it was displayed in the lower graphic). On the other hand, we considered instead of the condition in which $E^{2}\sqrt{E^{2} \theta^{2}-1} \gg \theta$ and, therefore, we obtained a linear behavior of such graphic which was shown in the upper graph in the right side.

In Fig \ref{fs}, a similar analysis could be done. In this sense, we have exhibited three configurations to Helmholtz free energy $F(\beta,\theta)$, considering the primitive temperature in the early universe. The top left exhibited a very slow variation of $F(\beta,\theta)$ when $\theta$ changed. Besides, we considered small variations of $\theta$, and the graphic exhibited a constant behavior, as shown in the bottom graphic. On the other hand, we rather regarded a situation where $E^{2}\sqrt{E^{2} \theta^{2}-1} \gg \theta$. Thereby, we had a linear behavior with a negative angular coefficient, though which was displayed in the top right.

Likewise, Figs. \ref{ss} and \ref{cc1} exhibited different behaviors of entropy $S(\theta)$ and heat capacity $C_{V}(\theta)$ for diverse values of $\theta$ analogously to what we did in the analysis accomplished for $\tilde{\alpha}(\theta)$ and $F(\theta)$ considering high temperature regime. Specifically, in Fig. \ref{ss} the upper graph on the left hand showed a variation of $S(\beta,\theta)$ when $\theta$ started to increase. Moreover, having regarded rather a situation where there existed the limit when $E^{2}\sqrt{E^{2} \theta^{2}-1} \gg \theta$, one possessed a linear behavior with a positive angular coefficient which was shown in the upper graph on the right hand. In addition, the bottom one exhibited how entropy $S(\beta,\theta)$ behaved for close values of $\theta$. Furthermore, in Fig. \ref{cc1}, the top left showed a variation of $C_{V}(\beta,\theta)$ when $\theta$ started to increase drastically and if one rather regarded a situation where there was the limit when $E^{2}\sqrt{E^{2} \theta^{2}-1} \gg \theta$, one would have a constant behavior which was shown in the top right. Besides, the bottom one revealed how heat capacity $C_{V}(\beta,\theta)$ behaved for close values of $\theta$.

Now, let us focus on the generalization of the Podolsky electrodynamics with the Lorentz-symmetry violation. Fig. \ref{lv} displayed the behavior of the black body radiation as a function of $\eta$ and $\nu$ for fixed values of $\theta$ and $D_{00}$ i.e., $\theta$ = 10 and $D_{00}=1$, in the generalized Podolsky electrodynamics. Next, we analyzed Fig. \ref{tlv}, which was the compilation of all thermodynamic functions in order to provide a concise discussion on this recent electrodynamics. Considering the plot to $\Bar{\alpha}(\eta)$ for huge values of $\eta$, we obtained a strong positive inclination of such curve, which differed from the pure Podolsky theory due to its increasing characteristic. A similar analysis was shown in $\Bar{S}(\eta)$, but for a different range of $\eta$ though i.e., $0\leq \eta \leq 100$. Additionally, it had the same behavior encountered in the plot of $S(\theta)$. In the same range of $\eta$, we verified that there existed a negative curve for Helmholtz free energy $\bar{F}(\eta)$ as well being in agreement with the usual Podolsky case. Next, for the case of heat capacity $\Bar{C}_{V}(\eta)$, it was displayed a linear increase with a positive angular coefficient when $\eta$ started to increase. Also, the behavior of this curve was completely different from heat capacity $C_{V}(\theta)$ exhibited in the Podolsky electrodynamics.


\section{Conclusion\label{conclusion}}

In this work, we studied the thermodynamic properties of a photon gas in a heat bath in the context of higher-derivative electrodynamics. We calculated the accessible states of the system in order to obtain the partition function that allowed us to investigate the behavior of the main thermodynamic functions, i.e., Helmholtz free energy, mean energy, entropy, and heat capacity. It is worth to be mentioned that all analyses were performed in the context of the primordial temperature of the universe, i.e., $\beta = 10^{-13}$ GeV. Moreover, we proposed a correction to the black body radiation in terms of the parameters $\theta$ and $\eta$, which accounted for Podolsky and its generalization with Lorentz violation, respectively.

Besides, we suggested corrections to the \textit{Stefan–Boltzmann} law for both theories. In the case of $\tilde{\alpha}(\theta)$, we saw that, depending on the considerations ascribed to the mean energy, we might find different behaviors of such curves. Likewise, Helmholtz $F(\theta)$, entropy $S(\theta)$, and heat capacity $C_{V}(\theta)$ could also have three different curves depending on the initial circumstances ascribed to them.

On the other hand, the parameter $\Bar{\alpha}(\eta)$ showed that there existed an accentuated increase as far as positive values of $\eta$ were taken into account. The Helmholtz free energy $\Bar{F}(\eta)$ started to decrease when $\eta$ increased. Both $\Bar{S}(\eta)$ and $\Bar{C}_{V}$ increased for positive changes of $\eta$, but the latter exhibited an increase with a constant angular coefficient though.

Finally, the physical consequences of the thermodynamic functions of such theories should be taken into account in order to possibly address some new fingerprints of a hidden experimental physics. Such analysis could be confronted by future experiments in the existence of Lorentz symmetry breaking. With this theoretical background, we can address a toy model for further investigations stepping towards to discover any trace of Lorentz violation. Furthermore, this thermal study might also be useful in future applications considering condensed matter physics for instance. Within the context of Lorentz violation, as future perspective, we could analyze the behavior of the thermodynamic properties for the generalized Podolsky with Lee-Wick terms. Additionally, knowing whether such theory is in agreement with either Cosmic Microwave Background regime or the electroweak epoch of the Universe seem also to be interesting investigations.


\section*{Acknowledgments}
\hspace{0.5cm}The authors would like to thank the Conselho Nacional de Desenvolvimento Cient\'{\i}fico e Tecnol\'{o}gico (CNPq) for financial support. R. V. Maluf  thanks CNPq grants 307556/2018-2 for supporting this project. A.A.A.F thanks D.M. Dantas for help with the computer calculations and L.L. Mesquita for the careful reading of this manuscript.


\end{document}